# Transmission-Aware Bandwidth Variable Transceiver Allocation in DWDM Optical Networks

Dmitry Khomchenko[1], Sai Kireet Patri[2,3], Achim Autenrieth[2], Carmen Mas-Machuca[3] and André Richter[4]

[1] *VPI Development Center,* Minsk, Belarus
*Email*: *dzmitry.khomchanka@vpi-minsk.com*

[2] *ADVA, Martinsried, Germany*
*Email: {SPatri, AAutenrieth}@adva.com*

[3] *Chair of Communication Networks, Department of Electrical and Computer Engineering,*
*Technical University of Munich (TUM), Germany*
*Email: cmas@tum.de*

[4] *VPIphotonics GmbH, Berlin, Germany*
*Email: andre.richter@vpiphotonics.com*

*Abstract* — **This paper addresses the transmission-aware transceiver allocation problem of flexible optical networks for a multi-period planning. The proposed approach aims at assigning the best configuration of bandwidth variable transceivers (BV-TRX) considering the amplifier noise and nonlinear channel interferences using the incoherent Extended Gaussian Noise (EGN) model. The proposed solution improves the network throughput and spectrum utilization in the early planning periods and allocates lower number of BV-TRXs in later periods in comparison to algorithms presented recently. A heuristic approach to regenerator placement has also been applied achieving up to 25% transceiver and 50% spectrum utilization savings in comparison to configurations without regenerators.**

***Keywords—network planning, elastic optical networks, flexible optical networks, transceiver allocation, traffic demand, routing, regenerator, OSNR, nonlinear interference, BVT, RSA, flex-grid, system throughput, EGN, GSNR***

## I. Introduction and Problem Definition

With the continuation of lockdown measures due to the COVID-19 pandemic, the use of digital communications at homes and businesses has undergone a revolutionary change [1]. With added supply chain restrictions, long haul network providers operating Dense Wavelength Division Multiplexing (DWDM) networks need to quickly find solutions to upgrade their networks to cope with the increasing traffic demands (TDs), while utilizing minimal resources. Simultaneously, the availability of bandwidth variable transceivers (BV-TRX) offers network planners to flexibly allocate resources without excessively overprovisioning the network [2]. However, this advantage increases the complexity of planning algorithms, which must cater for varying data-rate, channel bandwidth, modulation format and corresponding receiver sensitivity for each configuration of the BV-TRX. Furthermore, the allocation of flexible grid (flex-grid) channels, with a minimum spacing of 12.5 GHz, requires the calculation of the non-linear interference (NLI) penalty, every time channels are added, removed or reconfigured in the network [3].

Green-field network planning for flex-grid networks undertakes multiple inter-dependent steps [4], starting with, *a)* routing each TD in the network, *b)* assigning a set of transceivers to the TD and assigning each transceiver with a configuration, such that the traffic request is met, and *c)* assigning central channel frequencies (or wavelengths) to each transceiver in the network while maintaining wavelength continuity for every routed TD. Additionally, regenerator points (RPs) maybe placed to extend optical reach on longer routed TDs. In a BV-TRX enabled flex-grid long haul optical network, adding RPs can also lead to reduction in the overall number of deployed BV-TRXs, to meet the yearly traffic [5].

Fig. 1 shows an example flex-grid point-to-point link with add/drop nodes A, B and C. Each of these add/drop nodes consists of BV-TRXs, which can achieve multiple configurations: $\tau_1$ to $\tau_4$. Let us assume there exist two TDs between A and C, each having a traffic request of 600 Gbps, and one TD between B and C of 500 Gbps. Given the set of possible configurations, a configuration assignment algorithm may use the physical reach limitations and NLI calculation to find a relevant Quality of Transmission (QoT) metric like Generalized Optical Signal to Noise Ratio (GSNR), defined as the ratio of the channel launch power to the linear and non-linear interference inferred on the channel. In Fig. 1, TD 1 can be achieved by 2 BV-TRXs, each configured to $\tau_2$ and $\tau_3$ respectively, as shown by the blue channels. Similarly, TD 2 can be achieved by 3 BV-TRXs (green channels). Finally, assuming that the QoT metric is satisfied, TD 3 can be achieved with a single BV-TRX configured to $\tau_4$. In this way, the assignment of configuration to a BV-TRX is an important addition to pre-existing Routing and Spectrum Allocation algorithms. Such algorithms are also known as Routing, Configuration, and Spectrum Allocation (RCSA) algorithms.

Therefore, we exploit in this work the advantages of using RPs along with BV-TRXs in a flex-grid C+L-Band DWDM optical network, by proposing a novel BV-TRX allocation heuristic for the RCSA problem in such networks. Our results, applying a realistic traffic model [6] on a 17-node German network, show that using RPs not only allows operators to meet the yearly requested traffic, but also leads to a reduction of up to 25% of the total number of BV-TRXs, and savings of up to 50% in spectrum utilization, as compared to a solution without regeneration.

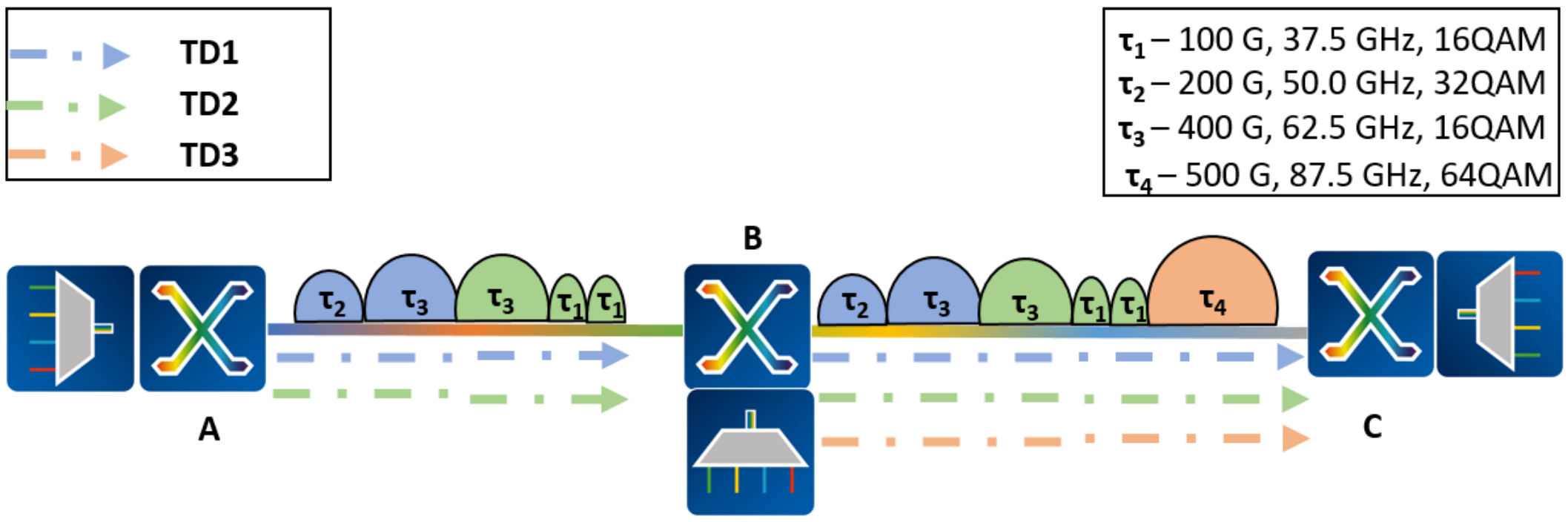

Fig. 1. An example network with 3 nodes, where 4 BV-TRX pairs assigned to 3 different TDs.

## II. Related Work

The BV-TRX allocation problem can be further divided into two sub-problems, namely, QoT calculation and configuration selection. Further, as discussed in the previous section, inserting RPs can help not only to increase the optical reach, but also enable BV-TRXs to achieve a higher network throughput. However, the decision to place RPs should be taken immediately after routing a TD, so that wavelength continuity can be ensured in the spectrum allocation step. Since our solution provides novelty in the BV-TRX allocation part of the larger RCSA workflow, we specifically survey three areas, namely, a) regenerator placement strategies, b) QoT calculation and c) configuration selection for BV-TRXs.

In flex-grid networks, regenerators are added by placing additional BV-TRXs at intermediate nodes to achieve higher end-to-end data rates. However, these regenerators need to be minimized to reduce additional costs incurred to operators. Since regenerator placement is an NP-hard problem, a comprehensive solution using genetic algorithm is presented in [7]. However, such algorithms have high computational and time complexity and simpler heuristics are often more desirable, which provide network planners quicker insights. The authors in [8] use a gradual RP deployment heuristic along network planning periods, to minimize the cost of placing RPs. The solution uses transmission reaches of different BV-TRX configurations, along with the requested demand throughput to place RPs in the network. However, higher data-rates such as 400 Gbps, which could significantly reduce the number of RPs, are not considered.

Different BV-TRX configurations have their own minimum receiver sensitivity Optical Signal to Noise Ratio (OSNR) defined as the minimum OSNR in decibels (dB), which needs to be achieved, such that the lightpath (LP) can be deployed. To find the QoT of each LP, its calculated GSNR should be always greater than the minimum OSNR of the corresponding receiver. The GSNR depends on both, the linear noise due to in-line amplifiers and NLI due to neighbouring channels. Previous works presented in [9, 10] use a Gaussian Noise Model based QoT estimation to find the GSNR for each LP. However, these values can change as new LPs are added to the network, which may lead to re-calculating the GSNR for all LPs. To avoid such an increase in computation, some improved QoT calculation strategies can be exploited. The authors in [11] explore some pre-computing strategies for traditional Routing and Spectrum Assignment (RSA) problems. A similar pre-computation of GSNR, in the first period of planning, provides simple bounds for subsequent configuration selection algorithms, while assuring that no LP falls below the minimum OSNR.

Several RCSA algorithms have been studied by the authors in [5, 6, 10] , each with different objective functions, based on certain assumptions. For example, authors in [5] assume that traffic of only 37.5 GHz is requested every time, which reduces the complexity of their configuration selection. However, in real-life deployment scenarios, the requested traffic that can be allocated to each BV-TRX can be anywhere between 25-100 GHz. This increase in generality also increases the solution space of BV-TRX deployment algorithms.

Thus, in a multi-period network planning scenario, all the algorithms such as routing, RP allocation, GSNR calculation, BV-TRX configuration generation, and spectrum assignment need to be used in a sequenced fashion to extract the network throughput and spectral efficiency. Such algorithms as routing and RP allocation can be applied only in the first planning period to reduce simulation time, the others should be executed in all periods due to traffic growth. The whole workflow of BV-TRX configuration in multi-period planning is discussed in the next section.

## III. BV-TRX Allocation Workflow

In this chapter, we introduce the optical network topology, along with related definitions and assumptions. Then, we present our proposed BV-TRX allocation workflow, which can be applied to a multi-period network planning scenario. The BV-TRX allocation workflow is divided into three major parts, namely, routing and regenerator placement (occurring once before the planning begins), GSNR calculation, and BV-TRX allocation (occurring every planning period). These parts are discussed in detail below.

### A. *Physical Network Topology and Impairements*

We consider an optical network topology as a set of nodes with add/drop capabilities. Some pairs of nodes are interconnected by links. Each link represents a trail of optical fiber spans, which differ in terms of length and type and are interconnected by in-line amplifiers with a gain being equivalent to the span attenuation.

As shown in Fig. 1, several TDs can share the same source-destination pair and each TD can be assigned multiple BV-TRX pairs, one each at the source and destination, respectively. For physical transmission impairments, we also distinguish between OSNR and GSNR. The former considers only the Amplified Spontaneous Emission (ASE) noise component of the in-line amplifiers, while the latter considers both, ASE as well as NLI noise components. To extend the reach of longer routes, lumped and distributed (Raman) optical amplifiers are considered by the algorithm when the OSNR is calculated. Calculating the OSNR before the GSNR allows filtering out the TRXs that do not guarantee the minimum OSNR and hence, avoiding unnecessary NLI calculations It is implied that for the same transmission path, OSNR is always greater than or equal to GSNR, and GSNR keeps degrading as more channels are added into the network [3, 12].

### B. *Routing and Regenerator Placement*

The overall RCSA algorithm for all planning periods is shown in Fig. 2. The first step of this algorithm is to route all TDs using Dijkstra’s shortest path. Routing needs to be performed in advance, not only to calculate the OSNR for each TD, but also to find the TDs which would need a regenerator point.

Once the path of each TD is found and the scenario of BV-TRX with RPs is considered, the required RPs are placed according to two reach-limiting constraints, namely the maximum number of spans in the route and the total route length. Both constraints are determined based on the network characteristics. For a given path, if at least one of two constraints is violated, an RP is placed according to the algorithm Alg. 1. The algorithm traverses the links of the path, calculating current distance from the start node and number of spans. If the distance and/or the number of spans exceeds the corresponding constraint, an RP is allocated to the node between the current and the previous link. Then, the current distance and number of spans are reinitialized and the algorithm searches for a next location of RP until the end of the route is reached. This RP allows replacing the long TD by a trail of several shorter TDs. This continues for each TD, till there are no further violations of the reach-limiting parameters.

```
if checkRoute(currentTD, Constraints)
    allocateRPs(currentTD, RPlist, Constraints)
    updateTDlist(TDlist, currentTD, RPlist)
end if

sub updateTDlist(TDlist, currentTD, RPlist)
    for each RP in RPlist
        create newTD
        TDlist.add(newTD)
    end for
    TDlist.remove(currentTD)
end sub

sub allocateRPs(currentTD, RPlist, Constraints)
    spanCount, distance = 0
    for each link in currentTD.route
        update(spanCount, distance)
        if spanCount or distance > Constraints
            RPlist.add(link.startNode)
            distance = link.length
            spanCount = link.spanCount
        end if
    end for
end sub
```

Alg.1. RP allocation algorithm.

As an example, let us compare Fig. 3 (a) and (d). In Fig. 3(a), we see a three-node point-to-point link, with two routed TDs (green dashed lines), between N3-N1 and N3-N2, respectively. Let us assume that both TDs request 400 Gbps data rate and that a maximum of three spans can be traversed, before regeneration is needed. TD N3-N2 is seen violating the reach parameter and is split into TD N3-N1 and TD N1-N2, each having a demand of 400 Gbps as shown in Fig. 3(d), by placing a regenerator at intermediate node N1. In this way, several long TDs in the network are split into shorter TDs. This also means that several shorter TDs may exist between any source and destination in the network. These TDs are grouped together for GSNR calculation and BV-TRX allocation, which occurs at every planning period.

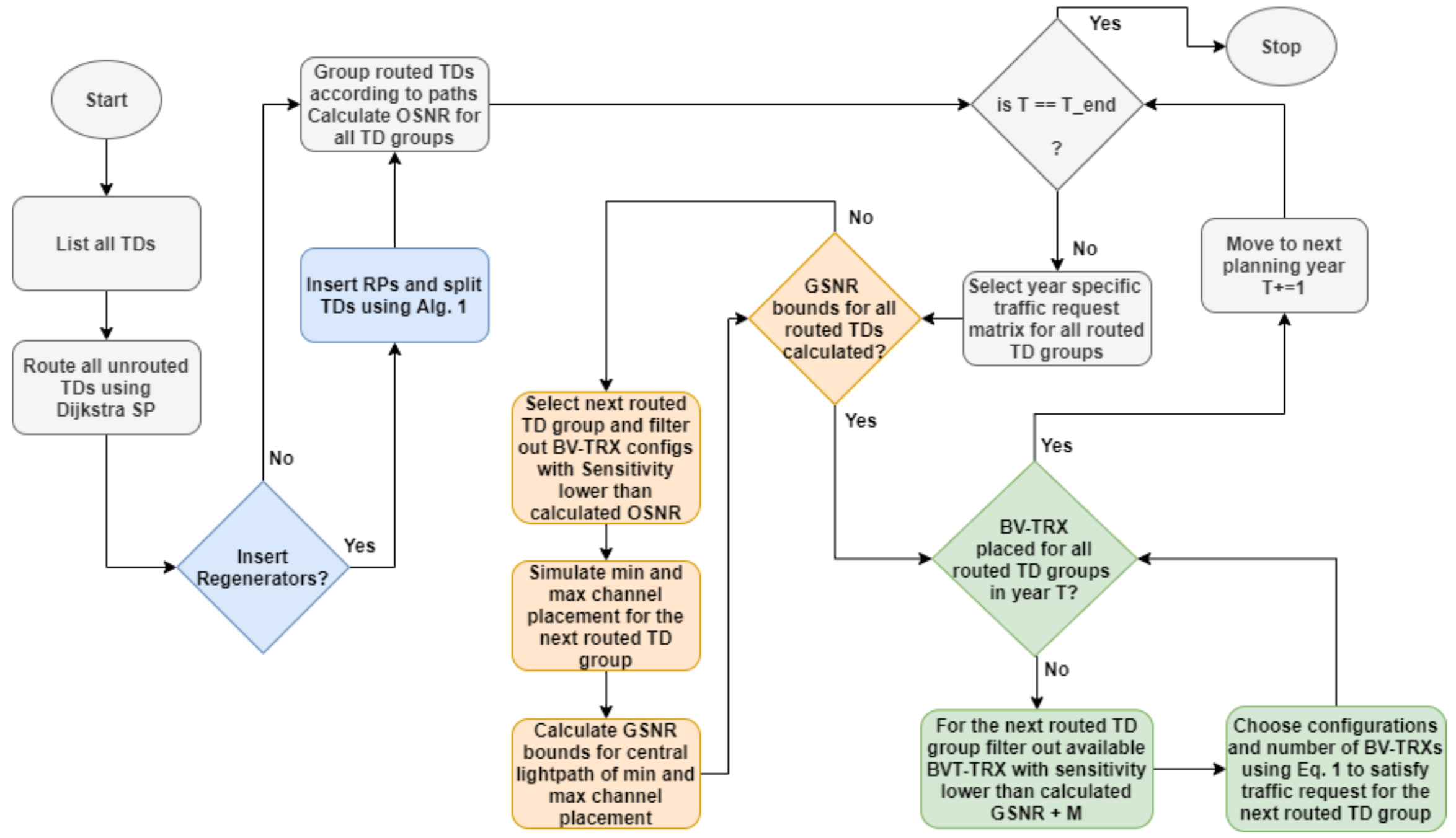

Fig. 2. BV-TRX allocation workflow diagram for all planning periods.

### C. *GSNR Calculation*

Calculating the GSNR is an important step to find the appropriate BV-TRX configuration, however, the exact channel conditions in a multi-period traffic scenario are unknown. This is because the NLI penalty calculations need the channel placement information along the transmission path of the BV-TRX. To avoid this race-around condition, we first group all TDs sharing the same source, destination, and route. By grouping TDs, we can sum all their requested data rate in each planning period into a group data rate. Firstly, we filter only those BV-TRX configurations, whose minimum OSNR is greater than the OSNR of the route of the grouped TDs. We then find two configurations with the minimum and maximum data rate. After this, we simulate two homogenous channel placements around the center of the C-Band of these configurations to meet the group data rate. For these two channel placements, we find the GSNR of the central channel based on the EGN model [12]. We only calculate the GSNR of the central channel, on the assumption that the central channel has the highest amount of cross channel interference and therefore the worst-case GSNR value. Therefore, we obtain two worst-case GSNR values from the minimum and maximum configuration placement and create a bound on the GSNR value for a particular group of TDs in the given planning period.

For example, as it is shown in Fig. 3(a) there are two TDs, TD N3-N1 and TD N3-N2, each requesting 400 Gbps at a particular planning period. Let us assume that according to OSNR degradation, BV-TRX configurations with data rate from 100G to 300G can be used for these TDs. For the first TD group (only TD N3-N1), we could satisfy the requested 400 Gbps, either by placing four 100 Gbps lightpaths or two 300 Gbps LPs. For both simulated placements, we calculate the GSNR only for the central channel. Let us assume that the value from the 300 Gbps homogenous placement is 16 dB and the value from the 100 Gbps homogenous placement is 12 dB. This means that for TD N3-N1, we can only place BV-TRX configurations with OSNR threshold greater than 16 dB. In another words, the worst GSNR will be used further to filter available BVT-TRXs. Similarly, we would get a different GSNR bound for the second grouped TD (TD N3-N2), since the distance traversed by the second group is different.

We note that although this calculation step may lead to deviations from the actual GSNR values, the accuracy can be sacrificed for faster computation time. In our method, the EGN model needs be run only twice per grouped TD per planning period. Such a method can enable planners to quickly run multiple studies and plan different solutions for the network under study.

### D. *BV-TRX Allocation*

After completing the GSNR calculation (orange boxes in Fig. 2) for the current planning period, we proceed to start the BV-TRX allocation. We begin by selecting the traffic request matrix of the current planning period, for all the routed TD groups. Then, for each routed TD group, a list of applicable BV-TRXs is created, whose GSNR satisfies the minimum OSNR, while keeping a small margin for future LPs additions (1 dB in our simulations). The available BV-TRX with the highest and the lowest data rate defines the lowest and highest number of LPs, respectively, that can be assigned to the TD. For example, one 400 Gbps LP or four 100 Gbps LPs can be used for a grouped TD of 400 Gbps, assuming that BV-TRXs are able to achieve either 100, 200, 300, or 400 Gbps on the given route. To find the exact configuration of BVT-TRX to be allocated to each grouped TD (for example, 2*100 + 1*200 Gbps), the optimization problem (1) is used, which minimizes the total bandwidth of BV-TRXs assigned to the grouped TD,

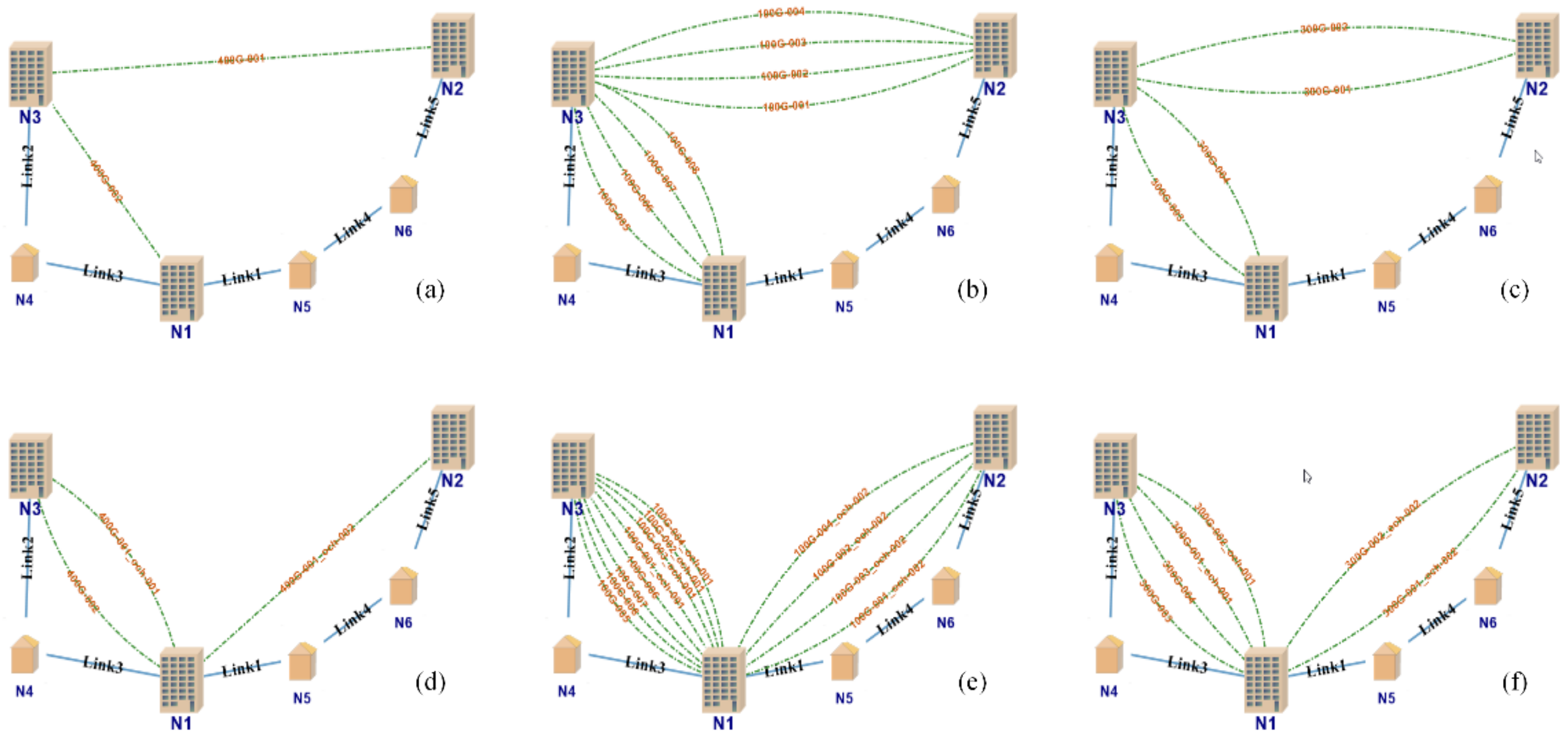

Fig. 3. Calculating upper and lower number of LPs for scenarios with and without RPs. (a) routed TDs without RP, (b) upper number of LPs without RP, (c) lower number of LPs without RP, (d) routed TDs with RP, (e) upper number of LPs with RP, (f) lower number of LPs with RP.

such that the sum of the data rates of all the TRXs is at least the TD data rate in every period.

$$
\begin{aligned}
&minimize\ \sum_i^N BW_i \\
&subject\ to\text{:}\ \sum_i^N BR_i \leq BR_{TD}\ , \qquad (1)\\
&\qquad GSNR_{TD} \geq T_{TRX,i} + M,\ \ \forall\, i \in N
\end{aligned}
$$

where $N$ is the number of LPs allocated to a grouped TD, $BW_i$ , $BR_i$ are the BV-TRX bandwidth and data rate, respectively, $BR_{TD}$ is the grouped TD data rate, $GSNR_{TD}$ is the lower bound of the GSNR values pre-calculated for a given grouped TD, $T_{TRX,i}$ is the BV-TRX minimum OSNR, and $M$ is the additional OSNR margin.

The minimization is done by comparing the bandwidth of all BV-TRX configurations. It is noted that the value of $N$ is a variable, based on the multiple configurations which meet the TD group's data rate. These are built through the use of an algorithm searching combinations with repetitions [13]. That is, all BV-TRXs configurations assigned to the TD will have their minimum OSNR higher than the worst-case GSNR, previously calculated. Then the configuration with the lowest spectral bandwidth will be selected as a solution.

In case that the workflow is unable to place any BVT setting for a routed TD, the TD is blocked. Afterwards, optical add/drop equipment is configured [14] and the system performance is assessed [15].

## IV. Simulation Results and Discussions

Since our workflow depends on OSNR and GSNR calculation, we first simulate a system as discussed in [16]. A value of 13 dB and 12 dB for OSNR and GSNR, respectively, is obtained, which is in good agreement with [16] and proves the validity of the implementation. For the comparison of system throughput, required TRX number and occupied frequency range, we consider a year as a planning period and undertake a ten-year planning (2021-2030) by creating a TD matrix for a 17-node German network based on the traffic growth model discussed in our previous work [6]. Fig. 4(a) shows the system throughput of our proposed heuristic (with and without RPs) and results from [6] with respect to the requested traffic. It can be observed that the solution with RPs is 15% closer to the requested overall traffic in the early planning years, as compared to the solution without RPs. Both are closer to the requested traffic than [6] in the early years, as shown in Fig.4(a). However, in the later years, i.e., 2027 onwards, both workflows tend to converge to the requested traffic.

The relative difference of simulation results of both algorithms for the scenarios without RPs during the ten-year planning is shown in Fig. 5. Our algorithm generates BV-TRX configurations that lead to effective spectrum utilization and system throughput in the first eight planning years. On the other hand, starting from 2026, the solutions are better in terms of the number of used BV-TRX.

As it is shown in Fig. 4(b, c) the total wavelength range occupied by all TRXs in a network is lower when RPs are used. Depending on topology and traffic (number of nodes, span length and requested system throughput) we found that the spectrum utilization can be two times lower in a solution with RPs, as shown in Fig 4(b, c). Fig. 4(b) shows w.r.t. the left axis the occupied spectrum range in terms of the equivalent number of wavelengths required for a conventional fixed grid of 50GHz. A higher number of TRXs along with a lower number of equivalent wavelengths in use compared to the algorithm [6] at early planning years (as shown in Fig. 4(b) and Fig. 5) is explained by the usage of different wavelength allocation algorithms.

Using our algorithm presented above more than 96 wavelengths are required in the last year of the ten-year planning without using RPs. Which means that for the last year, a C+L-band network should be built assuming the absence of dark fibers. However, the usage of RPs in the workflow allows planners to keep all wavelengths within the C-band for the whole planning. As shown in Fig. 4(b, c) the number TRXs allocated by the algorithm is lower for some combinations of network topology and TD matrices when RPs are used. This shows that along with an improved spectrum utilization, the usage of RPs may reduce the number of required TRXs for certain combinations of network topology and TD matrix. This effect can be explained by the grouping of routed TDs after inserting RPs and before allocating TRXs, as it is shown in Fig. 2. Besides, the usage of RPs splits the long paths in shorter ones reducing also the OSNR degradation. Hence TRXs with higher data rate and more advanced modulation formats can be used.

Similar simulations were performed for 14, 17 and 50 nodes networks of German topology [17] to compare the simulation times with the ones discussed in [9]. The TRX allocation workflow was tested on all three networks with and without RPs for different requested TD matrices using the workflow shown in Fig. 2. RPs were added using heuristic design constraints [15], i.e., RPs are added if parameters of original routed TDs (such as length and number of intermediate nodes) exceed certain limits. The simulation time with and without RPs measured for six planning years (2020-2025) is between 1 to 4 minutes for the 14- and 17-node networks, and between 90 to 120 minutes for the 50-node network, respectively.

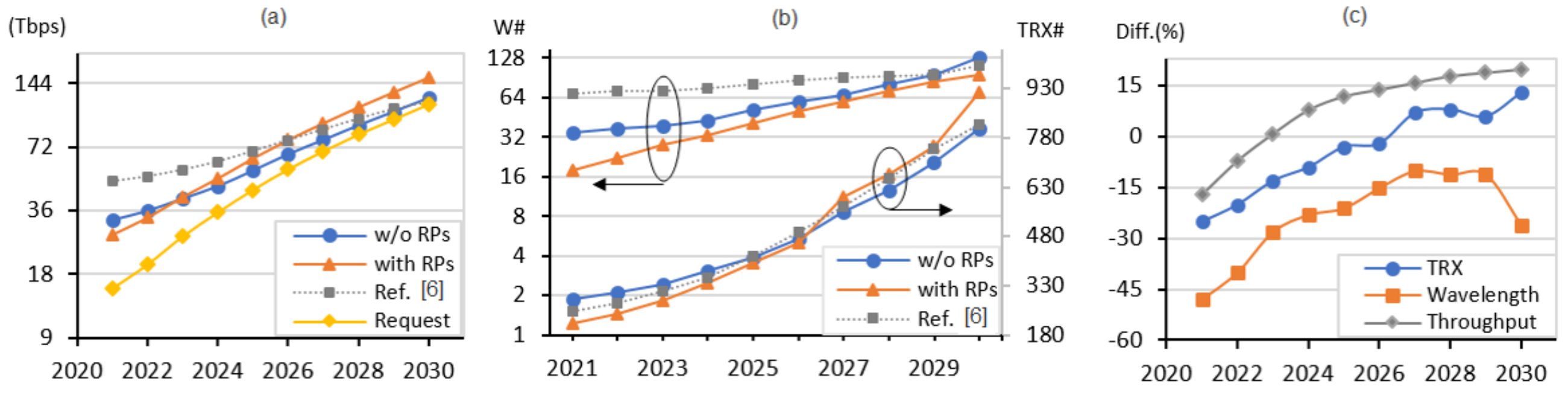

Fig. 4. Simulation results and comparison with Ref. [6]. (a) System throughput, (b) TRX and equivalent wavelength number calculated on 50 GHz fixed grid, (c) relative growth of system throughput, TRX and wavelength number in configuration with RPs with respect to configuration without RPs.

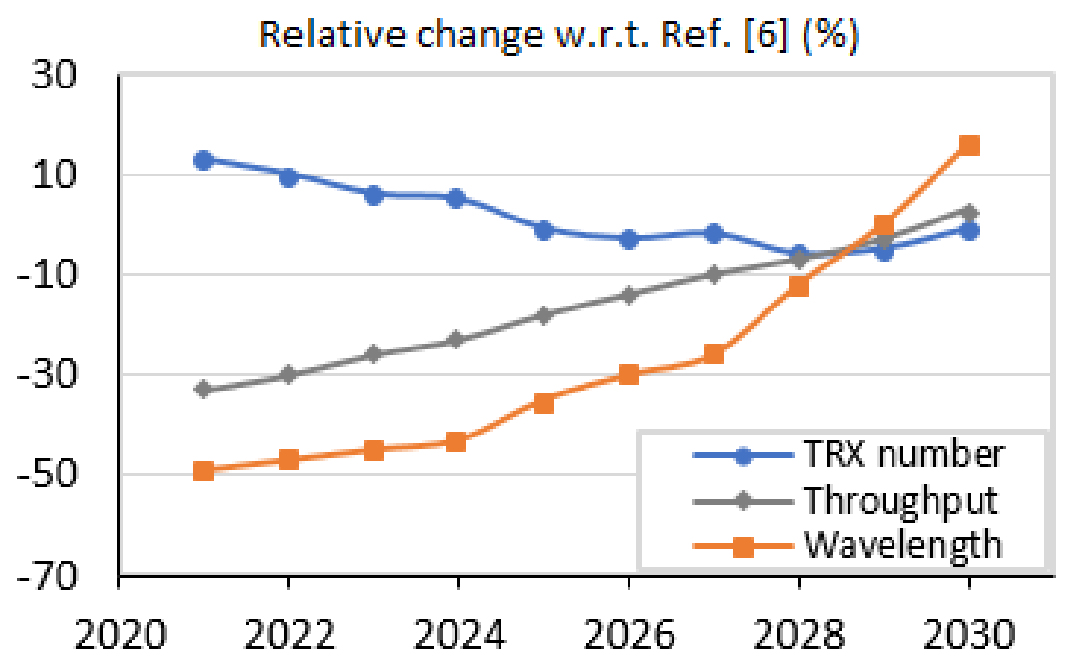

Fig. 5. Comparison of simulation results with Ref. [6]. Relative difference of system throughput, TRX and wavelength number in configuration without RPs.

Once transceivers and add/drop equipment have been configured, link loss compensation (LLC) is performed using the design environment of VPIlinkConfigurator [15]. LLC uses the same assumptions as in the TRX allocation workflow (exact span loss compensation). The algorithm allocated TRXs provide 1 dB OSNR margin as one of the constraints. The performance assessment after LLC, using methods discussed in [15], reports the minimum GSNR margin including NLI penalty of 1-1.5 dB for both configurations (with and without RPs). This confirms the feasibility of our approach.

## V. Conclusions

We presented a BV-TRX allocation workflow, which allocates BV-TRXs by accounting for linear impairments of the transmission line as well as for nonlinear interference penalties of neighboring optical channels. We showed that the usage of regeneration points allows improving spectrum utilization, to better control the overall network throughput, and for some network configurations to reduce the number of required transceivers.

## Acknowledgment

This work is partially funded by Germany's Federal Ministry of Education and Research (OptiCON, grant IDs #16KIS0993, #16KIS0989K, and #16KIS0991).